\documentclass[prb,showpacs,twocolumn]{revtex4}
\usepackage{graphicx}
\usepackage{amsmath}
\begin{document}

\let\vphi=\varphi

\newcommand\tr{\mathop{\textrm{Tr}}}
\newcommand\half{\frac{1}{2}}
\newcommand\lavg{\left\langle}
\newcommand\ravg{\right\rangle}
\newcommand\varE{{\mathcal E}}
\newcommand\im{\mathop{\textrm{Im}}\nolimits}
\newcommand\ket[1]{\left|#1\right\rangle}
\newcommand\bra[1]{\left\langle#1\right|}
\newcommand\braket[2]{\left.\left\langle#1\right|#2\right\rangle}

\title{Communicating Josephson Qubits}
\author{F. Plastina$^1$
    and G. Falci$^2$}
\affiliation{%
        $^{(1)}$ NEST-INFM $\&$ Scuola Normale Superiore, I-56127 Pisa, Italy\\
        $^{(2)}$ NEST-INFM $\&$ Dipartimento di Metodologie Fisiche e
        Chimiche (DMFCI),
        Universit\`a di Catania, viale A. Doria 6, I-95125 Catania,
        Italy}
\date{\today}

\begin{abstract}
We propose a scheme to implement a quantum information transfer
protocol with a  superconducting circuit and Josephson charge
qubits. The information exchange is mediated by an L-C resonator
used as a data bus. The main decoherence sources are analyzed in
detail.
\end{abstract}
\pacs{74.50.+r,03.67.Hk,73.23.Hk}

\maketitle One of the main purposes of quantum information
processing is the  faithful transmission of quantum states between
distant parties,  eventually exploiting entanglement among
subsystems. Examples include quantum teleportation
\cite{tele} and dense coding \cite{dense}, both of them
demonstrated using entangled photon pairs \cite{kn:ent-exps}, with
the ultimate aim of performing quantum cryptography \cite{crypto}.

Until now, however, much of the work has been done within the
realm of quantum optics\cite{pho} and very little efforts have
been devoted to describe and implement these phenomena with solid
state devices. On the other hand, nano-electronic devices have
been proposed as candidates for quantum computer implementation
\cite{dot,jj1,jj2,jj3} since they are easily embedded in
electronic circuits and scaled up to contain a large number of
qubits. In particular, superconducting Josephson junction
circuits, whose fabrication is now performed with well established
lithographyc methods, combine the intrinsic stability of the
superconducting phase with the possibility of controlling the
circuit dynamics through  manipulations of the applied voltages or
magnetic fluxes \cite{vion}. Direct experimental evidence that a
single-Cooper-pair box can be used as a controllable coherent two
level system  has been provided by Nakamura {\it et al}
\cite{nakamura}.

\begin{figure}\centering
\includegraphics[width=.75\linewidth]{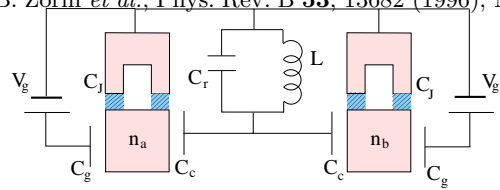}
\caption{Schematics of  the superconducting circuit with two
qubits coupled via the resonator.} \label{funo}
\end{figure}

Either the charge on the island or the phase differences at
junction can be used to store and manipulate quantum information
\cite{rmp}, the two regimes being characterized by dominating
charging and Josephson energies, respectively. Here, we
concentrate on the charge regime and propose a set-up that allows
 quantum information transfer and  entanglement generation between
two Josephson qubits. The circuit is designed so that interaction
between the two subsystems is mediated by an L-C resonator, see
Fig. \ref{funo}, playing the role of a data bus. The spirit of the
proposal is very similar to the Cirac-Zoller scheme for trapped
ion qubit \cite{ion}. As will be shown below, this set-up is
flexible enough to allow for quantum information transfer from one
qubit to another and for the generation of Bell states.
Furthermore, the circuit can be generalized to include more qubits
and we give the necessary prescriptions to  implement a universal
set of quantum gates.

The coupling of a single charge qubit to a large Josephson
junction (which may implement the resonator) has been  recently
exploited to perform on chip quantum state
measurements\cite{olivier}, and to prepare a mesoscopic
Schr\"odinger cat state~\cite{bruder}. We consider here a similar
coupling, but replace the single junction with a SQUID to achieve
a tuning of the Josephson energy~\cite{jj1}, which allows to
operate in both the dispersive and the resonant coupling regimes,
as required by the protocols described below.

We first analyze the model for a single qubit coupled to the resonator.
Letting $\vphi_J$ be the effective phase for the SQUID
\cite{rmp}, $\vphi_r$ the phase difference across the resonator
capacitance,
and $Q$ and $P$
their conjugated charges, the system Hamiltonian reads ($\hbar =1$)
\begin{eqnarray}
H&=& \frac{Q^2}{2 C_q} - \kappa_q V_g Q -
E_J(\Phi) \cos (2e \vphi_J) +
\frac{P Q}{C_{k}} \; \label{uno}  \nonumber \\
&&+ \frac{P^2}{2 C_{p}} + \frac{\vphi_r^2}{2 L} - \kappa_p V_g P
\end{eqnarray}
where the capacitances are given by (see Fig~\ref{funo}) $C_q= C_g
+ C_J + (C_c^{-1} + C_r^{-1})^{-1}$, $C_{p}= C_r + [C_c^{-1}+
(C_c+ C_J)^{-1}]^{-1}$, $C_{k}= C_{q} (C_c+ C_r) /C_c$, and with
attenuation parameters $\kappa_q=C_g/C_{q}$ and
$\kappa_p=C_g/C_{k}$. The relevant energy scales are the charging
energy, $E_{ch} = 2e^2/C_q$, the resonator frequency $\omega_r =
(L C_{p})^{-1/2}$, and the effective Josephson coupling of the
SQUID, $E_J (\Phi_x) = E_{J_0} \cos (2e \Phi_x)$, tunable via an
external magnetic flux $\Phi_x$.

We consider the charge regime ($E_{ch} \gg E_J$), where only the two lowest charge states
($Q=0,2e$) of the small island come into play, allowing to employ it as a
qubit. The electrostatic splitting between these two states is determined by the gate
voltage $V_g$, which we fix by setting $Q_g \equiv C_g V_g=e$.
This choice is crucial for what follows, as we will show that decoherence effects
are strongly quenched at this working point.
The eigenstates of the qubit Hamiltonian,
$ \ket{\pm} = ( \ket 0 \pm \ket{2e}) /\sqrt 2$, are then used as logical basis states.

If the qubit and
the oscillator are tuned near resonance, $\omega_r \sim E_J$,
the system described by Eq. (\ref{uno}) implements the Jaynes-Cummings
Hamiltonian \cite{jcm}
\begin{equation}
H_{JC}= \frac{E_j}{2} \sigma_z + \omega_r a^{\dag} a - i g
(\sigma_+ a - a^{\dag}  \sigma_-) \label{due}
\end{equation}
where $\sigma_z = \ket + \bra + - \ket - \bra- $, $ \sigma_+ =
(\sigma_-)^{\dag}= \ket + \bra -$ and $g = C_{k}^{-1} e \sqrt{2
\omega_r C_{p}}$. To obtain Eq. (\ref{due}), we introduced the usual ladder 
operators for the resonator and performed the Rotating Wave Approximation (RWA),
assuming the coupling to be  almost resonant and weak, $\delta=
E_j - \omega_r \ll E_j+ \omega_r$ and $g \ll \omega_r, E_j $. This
Hamiltonian generates Rabi oscillations between the states
$\ket{+,n_r}$ and $\ket{-, n_r+1}$ at the frequency $2 {\cal
R}_{n_r}= \sqrt{\delta^2 + 4 g^2 (n_r+1)}$. We will need to take
into account only oscillator states with at most $n_r=1$.
Exploiting the external flux dependence of $E_j$, it is possible
to switch between nearly resonant ($\delta \ll g$) and dispersive
regime ($g \ll  \delta \ll \omega_r$). In the latter case, the
time evolution is effectively  generated by
\begin{equation} H_{int}^{eff}=
\frac{g^2}{\delta} \left [ a a^{\dag} \ket + \bra + - a^{\dag}a
\ket -  \bra - \right ] \; .
\end{equation}
The resonant coupling allows to accomplish a quantum
state transfer, whereas switching between the two regimes is
required to perform a two--bit gate. If $\omega_r$ and $E_J$ are
very different from one another, the coupling  is effectively
switched off and the qubit evolves independently from the
resonator.

As suggested by \cite{olivier,bruder}, a large current biased
Josephson  junction can be used as a resonator. For bias current
$I$ well below the critical value $I_c$, the phase $\vphi_r$ of
the large junction is trapped in one of the minima of the tilted
washboard potential,
so that the system approximately behaves harmonically with
$\omega_r$  determined by the current.
This dependence of $\omega_r$ on $I$, gives a second
(and independent) mechanism to move to the resonant regime \cite{nota}.

We consider, now, the two--qubit set-up of Fig. (\ref{funo}), and
take for simplicity $C_c \ll C_{q}$, so that the direct
electrostatic interaction between the two qubit ($\sim
C_c^2/C_{q}^2$) can be neglected and they only interact through
the resonator via $H_{JC}$ (in fact, $C_{p}$, the $C_{q}$'s and
$g$ are slightly modified, but the  changes are negligible for
small $C_c$). To illustrate the use of the oscillator as a data
bus, we show how the quantum state of qubit $a$ can be transferred
to $b$. Let us suppose that the three subsystems are initialized
independently with qubit $b$ in $\ket -$ and the resonator in its
ground state:
\begin{equation}
|\psi,0\rangle = (c_+ |+\rangle \, + \, c_- |-\rangle) \otimes
|-\rangle \otimes |0\rangle \; .
\end{equation}
In the first step, the state of qubit $a$ is transferred to the
data bus by resonantly coupling them for a time $\tau=
\frac{\pi}{2 g}$. This leads to the state $\ket{\psi, \tau} = \ket
- \otimes \ket - \otimes (c_+\ket 1 + c_- \ket 0 ) \; .$ We then
de-couple qubit $a$ and perform the same  operation on qubit $b$.
Then, the system is led to
\begin{equation}
\ket{\psi, 2 \tau} = \ket - \otimes (c_+ \ket + + c_- \ket - ) \otimes \ket 0
\end{equation}
Thus, the state of one qubit has been transferred to  the other
one by exploiting the intermediary action of the resonator.

In a similar way, a maximally entangled singlet state can be
obtained by adapting a protocol already realized with atoms and
cavity\cite{hagley}. The underlying idea is very simple: first to
entangle $a$ and $r$, and then swap the entanglement by just
``exchanging" the states of the oscillator and qubit $b$. With the
system prepared in $\ket +_a \otimes \ket -_b \otimes \ket 0_r$,
we first let island $a$ and the resonator to interact resonantly
for a time $\tau/2=\pi/4g$ and then allow for the same coupling
(but lasting a time $\tau$) to be experienced by  island $b$. This
procedure gives rise to the EPR state $1/\sqrt{2}
\,(\ket{+-}-\ket{-+}) \otimes \ket 0 $. Note that, although the
oscillator is left in the ground state after the operations, it
actively mediates between the qubits. From the physical point of
view, this is the main difference with respect to the scheme of
Shnirman et. al \cite{jj1}, where the oscillator is only virtually
excited. As a consequence, when evaluating dephasing effects, the
oscillator needs to be included explicitly (as we will do below).

Besides quantum state transfer and entanglement generation, the
set-up allows to implement a universal set of quantum logic gates. 
Indeed,
single--bit rotations can be obtained by applying AC voltage pulses
on the qubit gate electrode. Furthermore, a two--bit gate
(equivalent to the control phase up to a one--bit operation) can
be accomplished through the following four steps {\it i)} couple
qubit $a$ to the oscillator in the dispersive regime for a time
$t_1$ (with $b$ de-coupled and the resonator initially prepared in
$\ket 0 $). This leaves the state $\ket{-}_a$ unaffected, while
appending the phase factor $e^{-i \theta}$ to $\ket{+}_a$, with
$\theta= \frac{g^2}{\delta} t_1$; {\it ii)} transfer the  state of
$a$ to the oscillator as in the previous protocol 
(i.e. let the two systems interact for a time $\tau$); 
{\it iii)} qubit $a$ being de-coupled, let $r$
and $b$ interact in the dispersive regime, again for time $t_1$;
{\it iv)} transfer back the state of the oscillator to qubit $a$
[same operation as in the step {\it ii})]. The resulting gate is
represented in the base $\{ |--\rangle, |-+\rangle, |+-\rangle,
|++\rangle \}$ as
\begin{equation}
\begin{pmatrix}
1 & 0& 0&0 \cr 0 & e^{-i \theta} & 0 & 0 \cr 0 & 0 & 1 & 0 \cr 0 & 0 & 0 & e^{i \theta}
\end{pmatrix} \; .
\label{sette}
\end{equation}

\begin{figure}
\centering
\includegraphics[width=.75\linewidth]{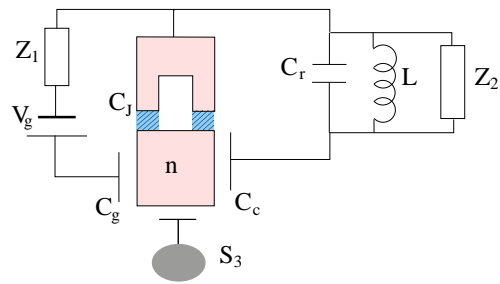}\
\caption{One qubit coupled to the resonator in the presence of the
decohering reservoirs (the third of which represents a bath of
fluctuating charge impurities).} \label{fdue}
\end{figure}

The treatment given so far has to be extended to account for
unwanted decoherence effect, whose major sources are
electromagnetic fluctuations of the circuit and noise originating
from bistable charged impurities located close to the islands. To
 estimate the time scales for relaxation and decoherence {\it
during} operations, we focus on the single--qubit plus resonator
scheme~\cite{michele}, depicted in Fig.~\ref{fdue}.

We first consider noise due to circuit impedances, modeled as
harmonic oscillators  reservoirs. Their effect on the system can
be described via the hamiltonian~\cite{CL,rmp}
\begin{eqnarray}
\delta H &=& \sum_{\alpha = 1}^{2} \Bigl [ H_\alpha^{env}
 - \hat{K}_\alpha  \, \hat{E}_\alpha \Bigr ]
+ B(Q,P,\vphi_r)
\label{eq:hamiltonian-environment}
\end{eqnarray}
Each $H_\alpha^{env}$ describes a set of harmonic oscillators. The coupling
term contains the operators $\hat{K}_\alpha$ acting on the system,
and collective environment operators $\hat{E}_\alpha$, whose
fluctuations determine decoherence.
The counter-term $B(Q,P,\vphi_r)$ enters only the proper determination of
energy shifts and will be disregarded from now on.

The explicit form of $\hat{K}_\alpha$ and $\hat{E}_\alpha$ can be obtained
by standard circuit analysis and by imposing that classical voltage and
current fluctuations are reproduced
in the proper limit~\cite{rmp,siewert}. This gives~\cite{longer}
$\hat{K}_1 = \kappa_{q} Q + \kappa_{p} P$ and $\hat{K}_2 = \vphi_r$,
and allows to identify
the fluctuation spectra $S_\alpha(\omega)$ of
the environment operators $\hat{E}_\alpha$
(i.e. the Fourier transforms  of their  symmetric equilibrium
correlation functions) as
\begin{eqnarray}
S_1(\omega) &=& \omega \; \mathrm{Re} \, \frac{Z_1(\omega)}{ 1 + i \omega Z_1(\omega)
C_{eff}(\omega)} \;
\coth \frac{\beta \omega}{2} \nonumber \\
S_2(\omega) &=& \omega \; \mathrm{Re} \, [Z_{2}(\omega)]^{-1} \;
\coth \frac{\beta  \omega}{2} \nonumber
\end{eqnarray}
with $C_{eff}(\omega) \simeq C_g$.

We now evaluate the effect of $\delta H$ on the eigenstates of $H_{JC}$
supposing a weak coupling with the environment (the
attenuation parameters $\kappa_q$ and $\kappa_p$ and the impedances can be
chosen to fulfill this condition).
The spectrum of $H_{JC}$
is made up of a ground state,
$\ket g=\ket{-,0}$, and a series of dressed doublets,
\begin{eqnarray}
&& \ket{a(n_r)} = \cos \theta_{n_r} \ket{+,n_r} + i \sin \theta_{n_r} \ket{-,n_r+1}
\quad \nonumber\\
&& \ket{b(n_r)} = i \sin \theta_{n_r} \ket{+,n_r} +  \cos \theta_{n_r} \ket{-,n_r+1}
\nonumber
\label{dress}
\end{eqnarray}
with eigenenergies $(n_r+ 1/2)\omega_r \pm {\cal R}_{n_r}$,
and where $\tan 2 \theta_{n_r}= 2 g\sqrt{n_r+1}/\delta$.
Only  $\ket g$ and the first doublet,
$\{\ket a , \ket b \}$, are involved in the coherent operations described so far.
In the secular approximation \cite{cohen}, relaxation and
dephasing rates in this subspace can be expressed in terms of the quantities
\begin{eqnarray}
\label{eq:funct-rates}
\gamma_{if}^\alpha(\omega) &=& 2 \, |\bra f \hat K_\alpha \ket i |^2 \;
S_\alpha (\omega) \; .
\end{eqnarray}
For instance, the dynamics of the populations is governed by a master equation
with transition rates given by
$\Gamma_{i \to f} = [1 + \mathrm{exp}(-\beta \omega_{if})]^{-1} \;
\sum_\alpha  \gamma_{if}^\alpha(\omega_{if})$, the standard Golden Rule result.

If $Z_2$ represents a resistor, the contributions of the second
bath become $ \propto 1/(Z_2 C_{p})$ (see table
\ref{table:coeff}), and simply reflect the finite quality factor
of the resonator. On the other hand, $Z_1$ affects both the qubit
and the oscillator, thus perturbing the overall system through two
interfering channels. As a result, the relaxation rate for $\ket
a$ is reduced if $\chi \kappa_p/\kappa_q\simeq 1$ for $\delta =0$.
Even if this condition is not met,
one of the eigenstates can be made more stable by choosing
an optimum $\delta$\cite{longer}.

\begin{table}[ht]
\begin{tabular}{l|cccc}
  &&$\ket i = \ket g$&$\ket i = \ket a$&$\ket i = \ket b$\\
\hline
$\alpha = 1$  &&    $e \kappa_q $&
    $- e (\kappa_q \, c - \chi \kappa_p \, s)$&
    $-i e (\kappa_q \, s + \chi \kappa_p  \, c)$
\\
$\alpha = 2$  && $0 $&$i (2 \chi e)^{-1} \, s$&
$(2 \chi e)^{-1} \, c$ \\
$\alpha = 3$  &&    $e C_q^{-1} $&
    $-e (C_{q}^{-1}  \, c - \chi C^{-1}_k  \, s)$&
    $ie (C_{q}^{-1}  \, s + \chi C^{-1}_k  \, c)$
\end{tabular}
\caption{Relevant matrix elements $\bra g \hat{K}_\alpha  \ket i$ of the
coupling operators with the electromagnetic ($\alpha=1,2$) and
the $1/f$ ($\alpha = 3$) environments.
Diagonal elements are equal, e.g. $\bra a \hat{K}_1 \ket a = \bra b \hat{K}_1 \ket b =
\bra g \hat{K}_1 \ket g = e \kappa_q$. Matrix elements
$\bra a \hat{K}_\alpha \ket b$ vanish. Here $c= \cos \theta_0$,
$s=\sin \theta_0$ and $\chi = \sqrt{C_p \omega_r/(2 e^2)}$.
}
\label{table:coeff}
\end{table}
Two important consequences come from the structure of the matrices
$\bra f \hat{K}_\alpha \ket i$ reported in
table~\ref{table:coeff}. First, all matrix elements between the
states of the doublet vanish, implying that the relatively small
frequency $\omega_{ab}$ never comes directly into play in the
rates. As a consequence, coherence is well preserved in the usual
temperature regime of operation, $ g < T \ll E_J $. A second, crucial
property is that each matrix $\bra f \hat{K}_\alpha \ket i$ has
equal diagonal elements. This implies that the dephasing rates,
$\Gamma_{ij}$, for the off-diagonal entries $\rho_{ij}$ of the
reduced density matrix, do not contain ``adiabatic'' terms, and,
therefore, are independent of the zero frequency spectra. Both
properties of the $K$'s directly result from the choice $Q_g=e$.
In a sense, the gate charge can be seen as a knob which allows to
operate at this ``optimal'' point, where low frequency noise does
not dephase the system. At the temperatures of interest the
largest dephasing rate is
\begin{equation}
\Gamma_{ab} = \frac{1}{2} (\Gamma_{a \to g} + \Gamma_{b \to g})
\approx \frac{1}{2} \sum_\alpha  [\gamma_{ag}^\alpha(\omega_{ag}) +
\gamma_{bg}^\alpha(\omega_{bg})]  \label{depharate}
\end{equation}
The quenched sensitivity to low frequency fluctuations is crucial
in the analysis of dephasing due to charged impurities lying close
to the island, responsible for $1/f$ noise.

Dephasing due to fluctuating impurities is believed to be the most
relevant problem in Josephson devices operating in the charge
regime. In general, for such an environment of fluctuators with a wide range
of switching rates, correlation times are too long for a master
equation approach to be always valid. Indeed, due to their
discrete character \cite{paladino}, slower fluctuators contribute
to decoherence in a distinctive manner, particularly marked when
adiabatic terms enter the dephasing rates. However, as shown
above,  dephasing due to small frequency fluctuations is minimized
at $Q_g = e$. In this case an estimate of the order of magnitude
of the effect can be obtained if the coupling with the environment
is treated to second order~\cite{napoli}, which is equivalent to
mimic the effect of fluctuating impurities with a suitable
oscillator environment. Then, Eqs.
(\ref{eq:funct-rates},\ref{depharate}) are still valid,
and the term describing $1/f$ noise contains
\begin{equation}
\label{kn:oneoverf}
S_3(\omega) = S_Q (\omega) = \frac{\pi A e^2}{\omega}
\end{equation}
where $S_Q (\omega)$ is the power spectrum of the charge fluctuations in the
island, whose amplitude can be inferred from independent
measurements~\cite{kn:exp}.

We now give some estimates of the relevant parameters of the
setup, and show that state transfer and entanglement generation
can be obtained with devices and circuits which are routinely
fabricated. For instance, we can take a large Josephson junction
as a resonator, with $C_r = 1 \, pF$, 
$\omega_r \simeq 37 \, \mu eV$. A low-temperature subgap resistance
$R_2 \gtrsim 600 \, K\Omega$ (here modeled by the parallel
impedance) can be achieved with Nb-based junctions, which
yields a quality factor $\omega_r R_2 C_r \gtrsim 4 \cdot 10^4$. For
the box we take $E_J = 40 \, \mu eV$ (eventually reduced by an
external flux), $C_J = 0.5 \, fF$, $C_g = 20 \, aF$ and $R_1 = 50
\, \Omega$. Furthermore, by taking $C_c =  50 \, aF$, we obtain
$g \simeq 0.5 \, GHz$ which allows operations on a time
scale $\lesssim 2 \, ns$. With this choice, the box charging
energy is $\simeq 0.6 \, meV$, so it operates in the  charging
regime. Moreover we have $g \ll \omega_r$ which ensures that the
RWA is valid, and, as we will see, $g$ is much larger than the
environment induced level broadening, which guarantees the
correctness of the secular approximation leading to equations
(\ref{eq:funct-rates},\ref{depharate}). These parameters
lead to the following estimates of the dephasing times due to
circuit fluctuations, $\tau_{\phi 1} \approx 1 \, \mu s$ and
$\tau_{\phi 2} \approx 1.20 \, \mu s$. For background charge noise,
$A=10^{-7}$ in equation (\ref{kn:oneoverf}) gives $\tau_{\phi
3} \approx 1 \, \mu s$. The resulting overall dephasing time is
$\tau_{\phi} = 1/\Gamma_{ab} \approx 376 \, ns$, allowing for the
two communication protocols. To achieve the two-bit gate, a somewhat 
larger dephasing time is required, which
can be obtained within the present technology by improving 
the quality factor of the resonator.

We thank B. Ruggiero, O. Buisson, M. Governale, Yu. Makhlin, E.
Paladino,  G. M. Palma and R. Fazio, F. Hekking for many helpful discussions.
This works has been supported by EU under IST-FET contracts EQUIP and SQUBIT,
by INFM under contract PRA-SSQI, and by MIUR under cofin 2001028294.

\begin{thebibliography}{10}
\bibitem{nota}
Since $I$ cannot become so large,
one necessarily has to change $E_J(\Phi)$ to decouple the qubit.
\end{thebibliography}
\end{document}